**5-20-10**

# Heisenberg-limited interferometry with pair coherent states and parity measurements

Christopher C. Gerry and Jihane Mimih

*Department of Physics and Astronomy*

*Lehman College, The City University of New York*

*Bronx, New York 10468-1589*

Abstract

After reviewing parity measurement based interferometry with twin-Fock states, which allows for super-sensitivity (Heisenberg-limited) and super-resolution, we consider interferometry with two different superpositions of twin-Fock states, namely two-mode squeezed vacuum states and pair coherent states. This study is motivated by the experimental challenge of producing twin-Fock states on opposite sides of a beam splitter. We find that input two-mode squeezed states, while allowing for Heisenberg-limited sensitivity, do not yield super-resolutions, whereas we that find both are possible with input pair coherent states.

PACS numbers: 42.50.St, 42.50.Dv, 42.50.Ex, 42.50.Lc

## I. Introduction

Over the past three decades there has been a concerted effort to find states and measurements schemes for detecting the small phase shifts that are expected from gravitational waves passing through optical interferometers such as those employed in the LIGO and Virgo projects [1]. One approach is to replace the first beam splitter of a Mach-Zehnder interferometer with a device that creates a maximally path entangled state containing $N$ photons, a state of the form

$$|\psi_N\rangle = \frac{1}{\sqrt{2}}\left(|N\rangle_a |0\rangle_b + e^{i\Phi_N}|0\rangle_a |N\rangle_b\right), \tag{1}$$

often called a N00N state [2]. If photon number parity measurements [3] are performed on one of the beams exiting the second beam splitter, Heisenberg-limit phase shift measurements, i.e. with uncertainty in the phase shift measurement given by $\Delta\varphi = 1/N$, independent of the phase shift $\varphi$, can be achieved. Furthermore, the N00N states result in super-resolved interference fringes in the expectation value of the parity operator. That is, we find [3b,c] with the parity operator given by $\hat{\Pi} = (-1)^{\hat{n}} = \exp(i\pi\hat{n})$ where $\hat{n}$ is the photon number operator of one of the output beams, that

$$\langle\hat{\Pi}\rangle = \begin{cases} (-1)^{N/2}\cos(N\varphi + \Phi_N), & N \text{ even}, \\ (-1)^{(N+1)/2}\sin(N\varphi + \Phi_N), & N \text{ odd}. \end{cases} \tag{2}$$

The expectation value oscillates with $N\varphi$ such that with $\varphi = 2\pi/\lambda$ there are $\lambda/N$ fringe spacings reduced over those for a single photon by a factor of $N$, $\lambda/N$ being the de Broglie wavelength of the photons. This reduction of the fringe spacing results in phase super-resolution. However, super-resolution and super-sensitivity, i.e. Heisenberg-limited sensitivity, are not the same thing as has recently been pointed out Resch *et al.* [4] who showed that it is possible to obtain super-resolution even with classical light, though to obtain super-sensitivity quantum light is required.

The difficulty of generating the required initial optical number state of large *N*, which would subsequently need to be manipulated into a N00N state, led us to consider [3] the prospect of instead using maximally entangled coherent states of the form

$$|\psi_\alpha\rangle = \mathcal{N}\left[|\alpha\rangle_a|0\rangle_b + e^{i\Phi}|0\rangle_a|\alpha e^{i\theta}\rangle_b\right] \tag{3}$$

from which we found that for small phase $\Delta\varphi \approx 1/\bar{N}$ where $\bar{N} = |\alpha|^2$ is the average photon number of the single mode coherent state. The maximally entangled coherent states are the continuous variable analogs of the N00N states. By expansion of the coherent states in the number basis it is easy to see that the maximally entangled coherent states are superpositions of the N00N states. The expectation value of the parity operator of the relevant output field mode, for the case where $\Phi = 0$ and $\theta = 3\pi/2$, is given by

$$\langle\hat{\Pi}\rangle = \frac{e^{-\bar{N}[1-\cos\varphi]}}{1+e^{-\bar{N}}}\cos\left[\bar{N}\sin\varphi\right]. \tag{4}$$

In the limit of small phase shift $\varphi$ and $\bar{N}$ large we find that $\langle\hat{\Pi}\rangle \approx \cos\left[\bar{N}\varphi\right]$ which does display super-resolution with de Broglie wavelength $\lambda/\bar{N}$.

The effectiveness of the N00N states for obtaining Heisenberg-limited precision for phase-shift measurements can be seen through a simple argument based on the heuristic number-phase uncertainty relation $\Delta N \Delta\varphi \simeq 1$. If we think of the distribution of photons in either of the modes, the uncertainty of the number of photons present, then the photon number uncertainty must be $\Delta N = N$, all *N* being in one mode or the other (the *total* number of photons involved has no uncertainty). Thus we have $\Delta\varphi \simeq 1/N$. Similar considerations apply for the maximally entangled coherent states except now in terms of the average photon number $\bar{N}$: $\Delta\varphi \simeq 1/\bar{N}$.

Another approach to Heisenberg-limited interferometry is that proposed by Holland and Burnett [5] in which twin Fock states $|N\rangle_a |N\rangle_b$ are simultaneously fed into the input ports of a 50:50 beam splitter. Assuming parity measurements are made on one output beam of a Mach-Zehnder interferometer (MZI), for small phase shifts one can show numerically that the phase shift uncertainty approaches the Heisenberg-limit $\Delta\varphi = 1/(2N)$ in the limit of large $N$ (the total number of photons passing through the interferometer is $2N$ ) [6]. But there is again the problem of producing the required input number states, especially for large *N*. In this case we would have to produce identical number states and simultaneously inject them into the first beam splitter of the MZI.

An alternative would be to consider superpositions of the twin Fock states of the form

$$|\psi\rangle = \sum_{N=0}^{\infty} C_N |N\rangle_a |N\rangle_b, \qquad \sum_{N=0}^{\infty} |C_N|^2 = 1, \tag{5}$$

from which we might expect that $\Delta\varphi \approx 1/(2\bar{N})$ where $\bar{N} = \sum_{N=0}^{\infty} N |C_N|^2$. Note that $2\bar{N}$ is the average photon number for both modes for a state of the form of Eq. (5). For future use, the joint photon number distribution for these states before the first beam splitter will be given by

$$P(n_1, n_2) = \left| \sum_{N=0}^{\infty} C_N \delta_{n_1,N} \delta_{n_2,N} \right|^2. \tag{6}$$

In this paper we shall examine two choices of states of the form of Eq. (5), namely the two-mode squeezed vacuum states (TMSVS) and the Pair coherent states (PCS). We find that both sets of states lead to Heisenberg-limited phase uncertainty, but that the PCS are more robust in the sense that the phase uncertainty clings to the Heisenberg limit much more closely for larger phase shifts than is the case for the TMSVS. Furthermore, we find that the TMSVS does not yield the desired super-resolution though super-resolution *is* present in the case of the PCS.

## II. Review of twin-Fock state approach with parity measurements

We begin with a brief review of the results obtained in our earlier paper [6] where we studied the use of parity measurements for interferometery with input twin-Fock states $\left|N\right\rangle_a \left|N\right\rangle_b$. The setup for this scheme is pictured in Fig. 1. The first beam splitter of the MZI is, for convenience, taken to be described by the beam transformations $\hat{a}' = \left(\hat{a}+\hat{b}\right)/\sqrt{2}$, $\hat{b}' = \left(\hat{b}-\hat{a}\right)/\sqrt{2}$, which means that the reflected wave does not pick up a $\pi/2$ phase shift. On the other hand, the second beam splitter is assumed to be one that *does* produce the $\pi/2$ phase shift on the reflected wave. This choice of beam splitter arrangement ensures that we obtain Heisenberg-limited in the for phase shifts in the vicinity of $\varphi = 0$ as appropriate for a search for small phase shifts.

For the twin-Fock input, the state just after the first beam splitter is [6, 7]

$$\left|\psi_{2N}\right\rangle = \sum_{k=0}^{N} A_k^N \left|2k\right\rangle_a \left|2N-2k\right\rangle_b , \tag{7}$$

where

$$A_k^N = \frac{1}{2^N}(-1)^{N-k}\left[\binom{2k}{k}\binom{2N-2k}{N-k}\right]^{1/2}. \tag{8}$$

For reasons that will become clear shortly, this state sometimes goes by the name "arcsine state". Picking up the phase shift, assumed to be in the *b* mode, this state becomes, just before the final beam splitter,

$$\left|\psi_{2N}(\varphi)\right\rangle = \sum_{k=0}^{N} e^{i\varphi(2N-2k)} A_k^N \left|2k\right\rangle_a \left|2N-2k\right\rangle_b. \tag{9}$$

After the second beam splitter, a parity measurement is assumed to take place in the *b* mode. The parity operator for this mode is $\hat{\Pi}_b = \exp(i\pi\hat{n}_b)$ where $\hat{n}_b$ is the photon number operator for the *b* mode. For the input twin-Fock states, the expectation value of the parity operator is

$$\left\langle\hat{\Pi}_b\right\rangle = P_N\left[\cos(2\varphi)\right] \tag{10}$$

where the $P_N[x]$ is a Legendre polynomial. In Fig. 2 we plot this expectation value against $\varphi$ and for two values of *N* and we see rapid oscillations with $\varphi$, more rapid with higher *N*. Thus the twin-Fock state approach leads to super-resolution. The uncertainty in the phase measurement is given by

$$\Delta\phi = \frac{(\Delta\Pi_b)}{\left|\partial\left\langle\hat{\Pi}_b\right\rangle/\partial\varphi\right|}, \tag{11}$$

where $\Delta\Pi_b = \sqrt{1-P_N^2\left[\cos(2\varphi)\right]}$. In Fig. 3 we plot the phase uncertainty against the total photon number $2N$ for two values of the phase shift, namely for $\varphi = 0.0001$ and for $\varphi = 0.05$. Also included are the corresponding standard quantum limit, $\Delta\varphi_{\mathrm{SQL}} = 1/\sqrt{2N}$, and the Heisenberg limit, $\Delta\varphi_{\mathrm{HL}} = 1/(2N)$, associated with the twin-Fock input states. The noise for the smaller phase shift gives results that track almost with the Heisenberg limit, becoming essentially exactly Heisenberg-limited for large enough *N*. But even the results for the larger phase shift track very close to the Heisenberg limit apart from certain photon numbers. In this sense, the twin-Fock state approach is fairly robust against phase shifts that may not be so small. Furthermore, Heisenberg-limited interferometry with twin-Fock states has been shown, by Meiser and Holland [8], to be robust against losses.

The effectiveness of twin-Fock input states for interferometry can be understood in part by looking at the joint photon number distribution of the states obtained by beam splitting, i.e. the states $\left|\psi_{2N}\right\rangle$. That distribution is given by

$$P\left(n_1, n_2\right) = \left|\sum_{k=0}^{N} A_k^N \delta_{n_1,2k} \delta_{n_2,2N-2k}\right|^2, \tag{12}$$

whose nonzero elements are given by

$$P\left(2k, 2N-2k\right) = \left(\frac{1}{2}\right)^{2N} \binom{2k}{k} \binom{2N-2k}{N-k}, \qquad k \in \left[0, N\right], \tag{13}$$

and form a distribution known in probability theory as the fixed-multiplicative discrete arcsine law of order *N* [9]. In Fig. 4 we plot this distribution for the case $N = 10$. Note that the distribution is concentrated along an "anti-diagonal" line. The corresponding distribution for the N00N state differs in that *only* the extreme states $\left|20\right\rangle_a \left|0\right\rangle_b$ and $\left|0\right\rangle_a \left|20\right\rangle_b$ are populated, though they are also the most populated states for the arcsine states, there being a low plateau between these extremes.

## III. Two-mode squeezed vacuum states

For input states given by the superposition of Eq.(5), the state after the beam splitter is given by

$$\left|\psi_{\text{BS}}\right\rangle = \sum_{N=0}^{\infty} C_N \left|\psi_{2N}\right\rangle \tag{14}$$

with $\left|\psi_{2N}\right\rangle$ given by Eq. (7), and the expectation value of the parity operator becomes

$$\left\langle\hat{\Pi}_b\right\rangle = \sum_{N=0}^{\infty} \left|C_N\right|^2 P_N\left[\cos\left(2\varphi\right)\right]. \tag{15}$$

and the joint photon number distribution after the first beam splitter is given by

$$P\left(n_1, n_2\right) = \left|\sum_{N=0}^{\infty}\sum_{k=0}^{N} C_N A_k^N \delta_{n_1,2k}\delta_{n_2,2N-2k}\right|^2 \tag{16}$$

We first consider the TMSVS given by

$$\left|\xi\right\rangle_{ab} = \left(1-\left|\xi\right|^2\right)^{1/2} \sum_{N=0}^{\infty} \xi^N \left|N\right\rangle_a \left|N\right\rangle_b, \tag{17}$$

where $\xi$ is a complex number constrained to be $\left|\xi\right| < 1$. Obviously, $C_N = \left(1-\left|\xi\right|^2\right)^{1/2} \xi^N$. Such states are routinely produced via parametric down-conversion experiments [10]. In Fig. 5 we plot the expectation value of the parity operator with these coefficients as a function of $\varphi$. We see no oscillations at all with respect to $\varphi$, just a central peak at $\varphi = 0$, though the width of the peak scales as $1/\left(2N\right)$. In contrast to the case of the twin Fcok states, here *we do not observe super-resolution.* In Fig. 6 we plot the phase uncertainty against the total average photon number of the two modes, $2\bar{N} = \left\langle\hat{n}_a + \hat{n}_b\right\rangle = 2\left|\xi\right|^2 / \left(1-\left|\xi\right|^2\right)$ for the cases $\varphi = 10^{-4}$ and $\varphi = 0.05$ along with the corresponding standard quantum limit, $\Delta\varphi_{\text{SQL}} = 1/\sqrt{2\bar{N}}$, and the Heisenberg limit,

$\Delta\varphi_{\mathrm{HL}} = 1/(2\bar{N})$. We find that the phase uncertainty can be a bit below the Heisenberg limit though the effect is most noticeable for low average photon numbers. This was noticed and explained in terms of the Fisher information by Anisimov *et al.* [11]. But we notice a significant difference between the phase uncertainty results of this case and the case of the twin-Fock states. For the TMSVS the phase uncertainty for $\varphi = 0.05$ is close to Heisenberg-limited only for very small average photon numbers. Even for the case $\varphi = 10^{-4}$ we can see that the phase uncertainty starts to go up as we reach large average photon numbers. Thus overall it does not appear that the TMSVS state is optimal for interferometry.

The reasons for this, we believe, are two-fold. The joint photon number distribution of the TMSVS before and after beam splitting is given in Fig.7. The distribution prior to beam splitting is thermal-like down the diagonal as can be seen from the inset. It is broad, in fact, it is super-Poissonian in both modes, and is peaked for the vacuum states $|0\rangle_a|0\rangle_b$ instead of a twin-Fock state of high photon number. The joint distribution of this state after beam splitting clearly reflects the distribution before. The second reason is that the TMSVS taken as a whole, i.e. by not truncating the state as is done in perturbative approaches, becomes *disentangled* by the action of the beam splitter [12]. In fact, the state is transformed into a product of single-mode squeezed vacuum states $|\pm\xi\rangle$, i.e. the product state $|\xi\rangle_a|-\xi\rangle_b$ where

$$|\pm\xi\rangle = \left(1-|\xi|^2\right)^{1/4} \sum_{m=0}^{\infty} (\pm 1)^m \frac{\sqrt{(2m)!}}{2^m m!} \xi^m |2m\rangle. \tag{18}$$

That the superposition state of Eq. (14) for the coefficients of TMSVS is equivalent to the product state $\left|\xi\right\rangle_a \left|-\xi\right\rangle_b$ can be seen by expanding out the latter using Eqs. (18). Of course, the modes become re-entangled by the action of the second beam splitter. But, in contrast to the case of the twin-Fock states, and for that matter the N00N state, there is no entanglement inside the interferometer during phase encoding. The TMSVS is a Gaussian state and for that reason does not exhibit the degree of non-classicality of the N00N states, the maximally entangled coherent states, and the arcsine states.

**IV. Pair coherent states**

The pair coherent states (PCS) [13], also referred to as circle states [14], have the form

$$\left|\zeta\right\rangle = \mathcal{N}\int_0^{2\pi} \left|\sqrt{\zeta}e^{i\theta}\right\rangle_a \left|\sqrt{\zeta}e^{-i\theta}\right\rangle_b d\theta \tag{19}$$

where the $\left|\sqrt{\zeta}e^{\pm i\theta}\right\rangle$ are Glauber coherent states. In terms of the number state bases the PCS may be written as

$$\left|\zeta\right\rangle = \mathcal{N}_0 \sum_{N=0}^{\infty} \frac{\zeta^N}{N!} \left|N\right\rangle_a \left|N\right\rangle_b , \tag{20}$$

where the normalization factor $\mathcal{N}_0 = 1\big/\sqrt{I_0\left(2\left|\zeta\right|\right)}$ where $I_0\left(2\left|\zeta\right|\right)$ is the modified Bessel function of order zero. The parameter $\zeta$ is a complex number and the states satisfy the relations eigenvalue conditions $\hat{a}\hat{b}\left|\zeta\right\rangle = \zeta\left|\zeta\right\rangle$ and $\left(\hat{a}^\dagger\hat{a} - \hat{b}^\dagger\hat{b}\right)\left|\zeta\right\rangle = 0.$ The first shows that they are similar to the usual single-mode coherent states in that they are eigenstates of a lowering operator (actually a product of such operators), while the second says that the difference in the photon numbers of each of the modes must be zero, a condition that restricts the state to be a superposition of twin Fock states. The PCS were first discussed in the quantum optics literature by Agarwal [13] and it has recently been shown theoretically that such states could be produced by non-degenerate parametric oscillators [14]. The states have been discussed with many applications in mind, such as violations of EPR-Bell-type inequalities [15], and continuous variable quantum information processing such as quantum teleportation [16], quantum communication [17], and quantum cryptography [18] though in the last two references the PCS are instead called two-mode coherently correlated states, or states of correlated twin laser beams. Unlike the TMSVS, the PCS are non-Gaussian states. As far as we are aware, the present work constitutes the first application of the PCS to quantum metrology.

Using the coefficients $C_N = \mathcal{N}_0 \zeta^N / N!$ in our expression for the parity operator in Eq. (15) we can numerically determine the expectation value of the parity operator of the output *b* mode, which we plot as a function of the phase shift in Fig. 8, and the phase uncertainty the phase uncertainty against $2\bar{N}$, where $\bar{N} = \sum_{N=0}^{\infty} N |C_N|^2 = (\mathcal{N}_0/\mathcal{N}_1)^2 |\zeta|^2$ and where $\mathcal{N}_1 = \left[|\zeta^{-1}| I_1(2|\zeta|)\right]^{-1/2}$, for $\varphi = 10^{-4}$ and $\varphi = 0.05$ is given in Fig. 9, where we include for comparison $\Delta\varphi_{\text{SQL}} = 1/\sqrt{2\bar{N}}$ and $\Delta\varphi_{\text{HL}} = 1/(2\bar{N})$ For the smaller $\varphi$ we have sensitivity essentially at the HL whereas even for the larger value, apart from the large deviations for certain average photon numbers, the phase uncertainty is still fairly close to the Heisenberg limit. The PCS appear to be more robust for parity based interferometry than do the two-mode squeezed vacuum states. In Fig. 10 we present the joint probability distributions of the PCS both before and after beam splitting. In contrast to the case of the TMSVS, the photon number distribution for the PCS is highly peaked in the vicinity of $\bar{N}$ for each mode. Furthermore, the distribution is rather narrow; in fact, it is sub-Poissonian in both modes and thus can be understood as highly selective for the twin-Fock state $\left|N\right\rangle_a \left|N\right\rangle_b$ for $N \sim \bar{N}$. Sub-Poissonian statistics are nonclassical. The joint distribution after the first beam splitter clearly reflects that of the input state in that we see that, in contrast to the case of the TMSVS, there is a prominent ridge across the anti-diagonal indicating that the distribution is highly selective for arcsine states associated with large average photon number. The corresponding distribution for the TMSVS is highly selective only in the vicinity of the vacuum as we have already pointed out.

**V. Signal-to-noise ratio**

Another important consideration, along with super-resolution and super-sensitivity, is the signal-to-noise ratio (SNR). Some time ago, Kim *et al*. [19] considered the operator $\left(\hat{a}^\dagger\hat{a} - \hat{b}^\dagger\hat{b}\right)^2$ at the output of the interferometer as the measure for the phase shift. Though it leads to Heisenberg-limit sensitivities, it has a modest SNR of $\sqrt{2}$. In contrasts, the measurement of parity can have a very high SNR. For this measure, the SNR is defined as

$$\text{SNR} = \left\langle \hat{\Pi}_b \right\rangle \Big/ \Delta\,\Pi_b\,. \tag{21}$$

In Fig. 11 we plot for the input twin-Fock states, the TMSVS, and the PCS, the $\text{Log}_{10}\left(\text{SNR}\right)$ SNR versus $2\bar{N}$ (or *2N* in the case of the twin-Fock states). The SNR ration for all three is quite high with the twin-Fock and PCS cases being essentially identical.

## VI. Concluding remarks

In this paper we have examined the prospect of performing parity-measurement based Heisenberg-limited interferometry with superpositions of twin-Fock states, the twin-Fock states themselves known to reach such sensitivity in the limit of large photon number. Actually, the number of photons need not be all that large, but producing identical Fock states at the inputs of a beam splitter is a challenge, a challenge that could obviated by instead using a well-chosen superposition of twin-Fock states. The obvious case to try is the TMSVS, but, as we have seen, because its photon number distribution is thermal-like, this state is dominated by the twin vacuum state. Furthermore, it becomes disentangled by the action of the beam splitter, and while it leads to Heisenberg-limited phase uncertainty and high signal-to-noise ratio, it does not lead to super-resolution. Finally, even for small phase shifts, the phase uncertainty moves away from the Heisenberg limit for increasing average photon numbers, moving away quite rapidly for increasing phase shifts. But with the PCS we have a sub-Poissonian joint photon number distribution peaked near the average photon number in the two modes. This allows for Heisenberg-limited phase uncertainty, super-resolution, and high signal to noise ratio, results that are very close to those obtained from input twin-Fock states.

Finally, we have mentioned that pair coherent states have yet to be generated in the laboratory as far as we are aware, but that it has recently been shown theoretically that such states could be produced by non-degenerate optical parametric oscillators [14]. It is not unreasonable expect that within a few years pair coherent states will be available be available in the laboratory, and that the application described in this paper serving as a new motivation to press for their production.

**ACKNOWLEDGEMENTS**

We acknowledge support from the Department of the Army, The Research Corporation, and grants from PSC-CUNY.

**Figure Captions**

Fig. 1 Schematic diagram of a Mach-Zehnder interferometer with twin-Fock state inputs.

Fig. 2 Plot of expectation value of parity versus $\varphi$ for input twin-Fock states with $2N = 4$ (solid line) and $2N = 30$ (dotted line).

Fig. 3 Phase uncertainty against total photon number $2N$ for twin-Fock states with $\varphi = 10^{-4}$ (squares) and $\varphi = 0.05$ (dots). Included for comparison is the standard quantum limit $1/\sqrt{2N}$ (dot-dashed line) and the Heisenberg limit $1/(2N)$ dotted line).

Fig. 4 Joint photon number distribution for the arcsine state with $N = 10$.

Fig. 5 Plot of expectation value of parity versus $\varphi$ for input TMSVS with $2\overline{N} = 4$ (solid line) and $2\overline{N} = 30$ (dotted line).

Fig. 6 Phase uncertainty versus $2\overline{N}$ for the TMSVS with the same angles as before. Note that the curve for even the smaller angle $\varphi = 10^{-4}$ (donated by the squares) lifts away from the Heisenberg limit for large enough $2\overline{N}$. Though it is not easy to see because the effect is small, the phase uncertainty is actually below the Heisenberg limit for small values of $2\overline{N}$. This effect, which is very small, was pointed out in Ref. [11].

Fig. 7 Joint photon number distributions for the TMSVS (a) before the beam splitter and (b) after the beam splitter for $\overline{N} = 10$.

Fig. 8 Plot of expectation value of parity versus $\varphi$ for input PCS with $2\overline{N} = 4$ (solid line) and $2\overline{N} = 30$ (dotted line).

Fig. 9 Phase uncertainty versus $2\overline{N}$ for the PCS with the same phase shift angles as before.

Fig. 10 Joint photon number distributions for the PCS (a) before beam splitting and (b) after beam splitting.

Fig. 11 $\mathrm{Log}_{10}\left(\mathrm{SNR}\right)$ as a function of $2N$ or $2\bar{N}$ for the twin-Fock states, TMSVS and the PCS. The solid and dot-dashed lines that are essentially indistinguishable of for the twin-Fock states and PCS, respectively, while the dashed line is for the TMSVS.

Fig.1:

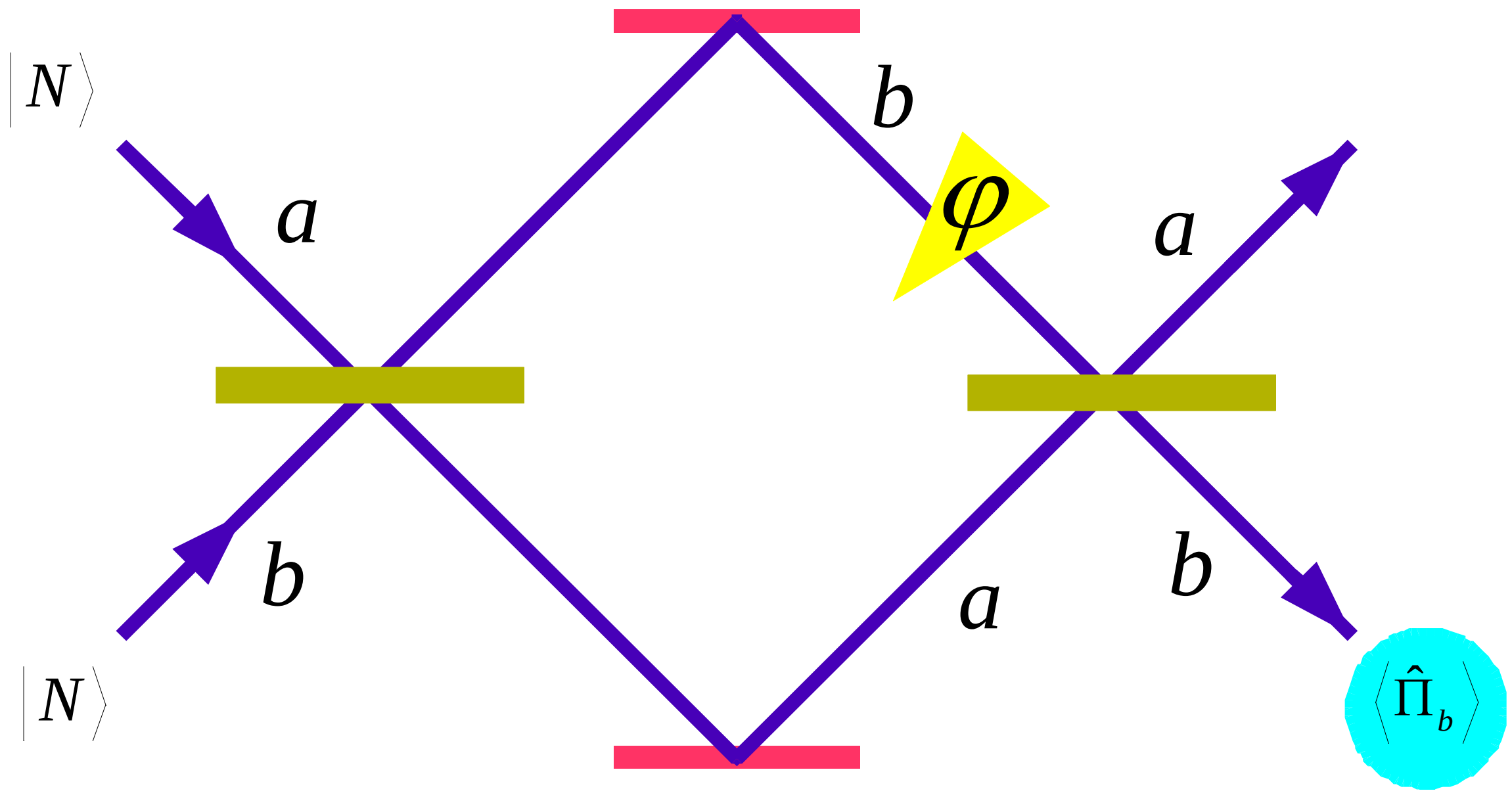

Fig. 2:

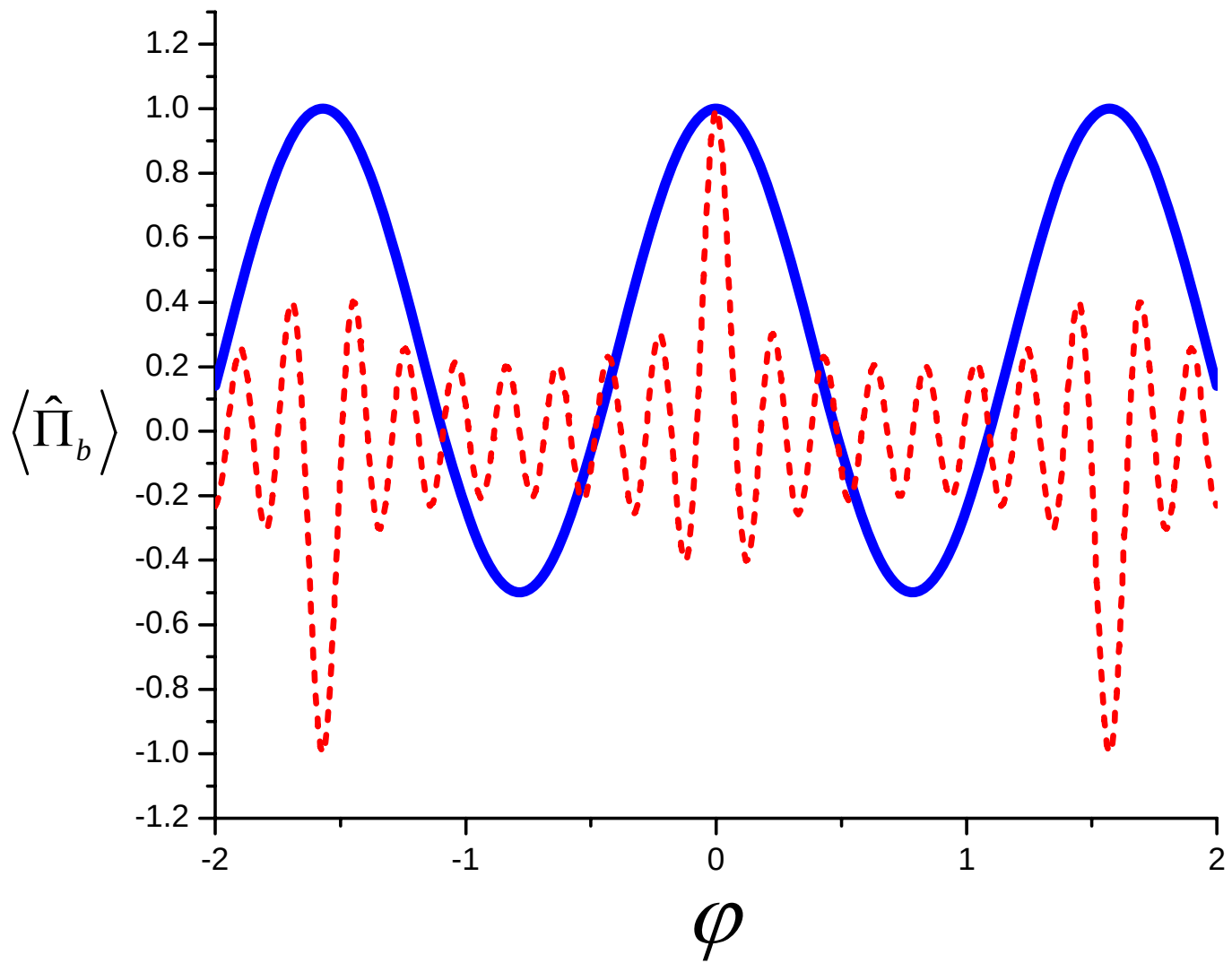

Fig. 3:

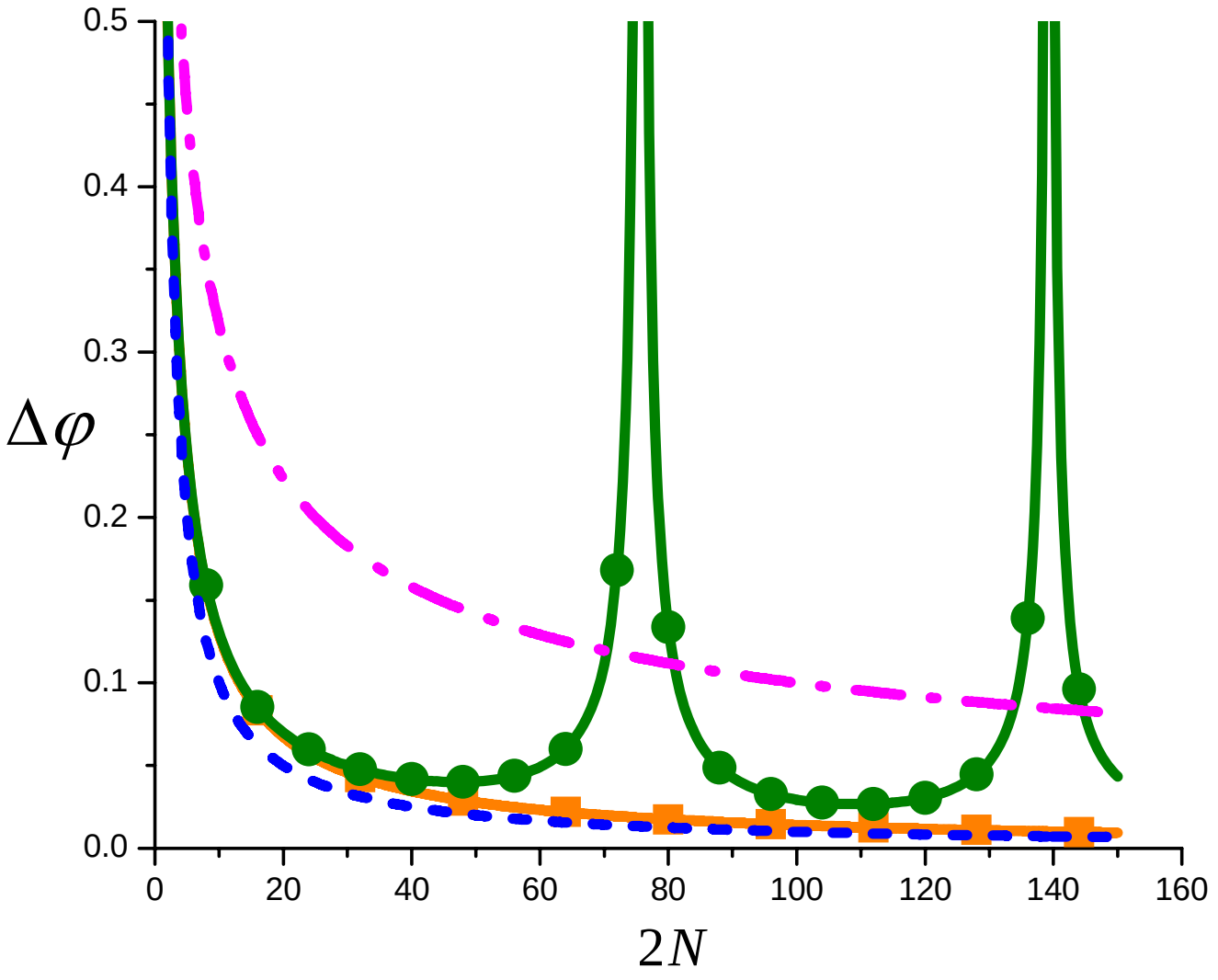

Fig. 4:

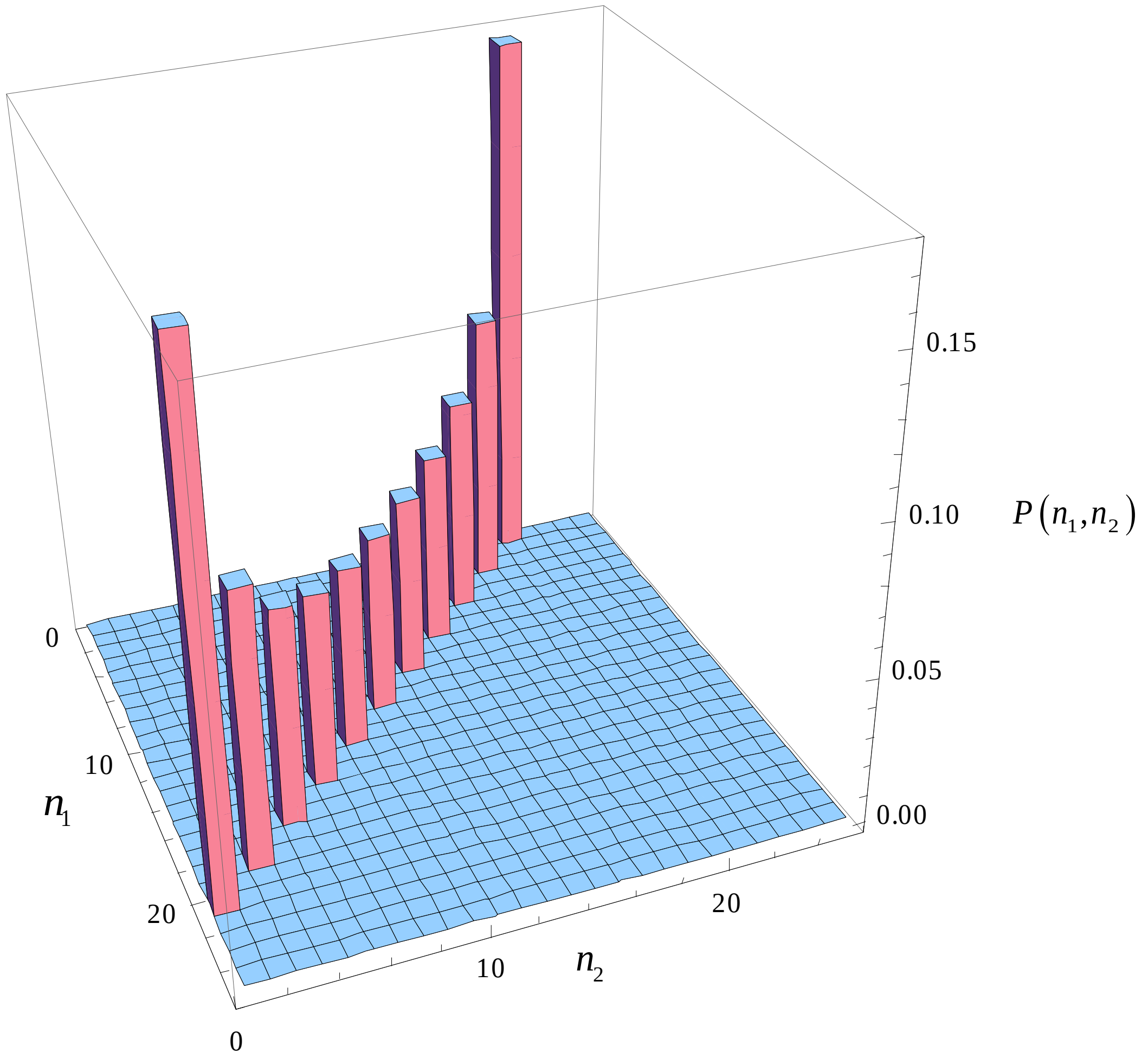

Fig. 5:

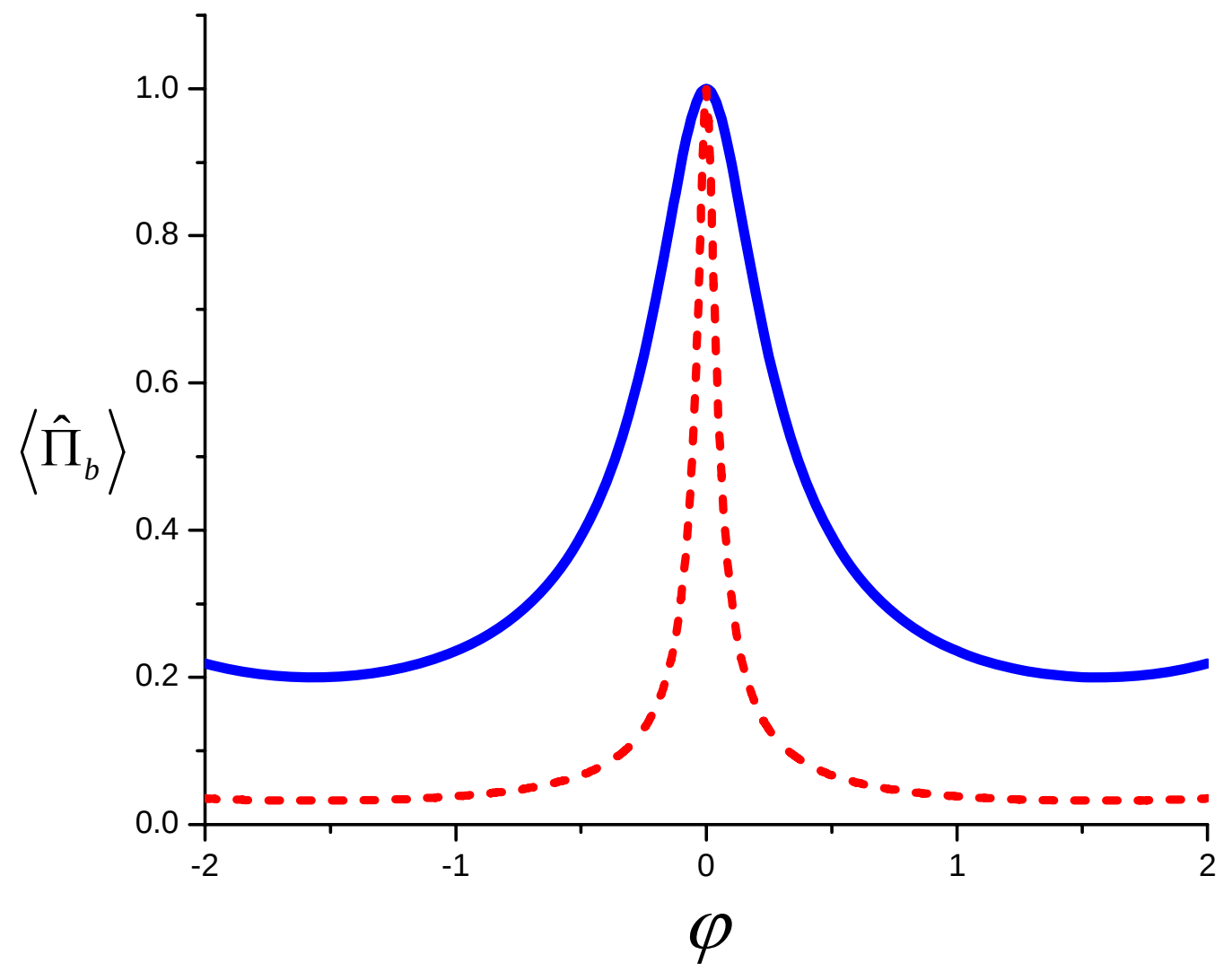

Fig. 6:

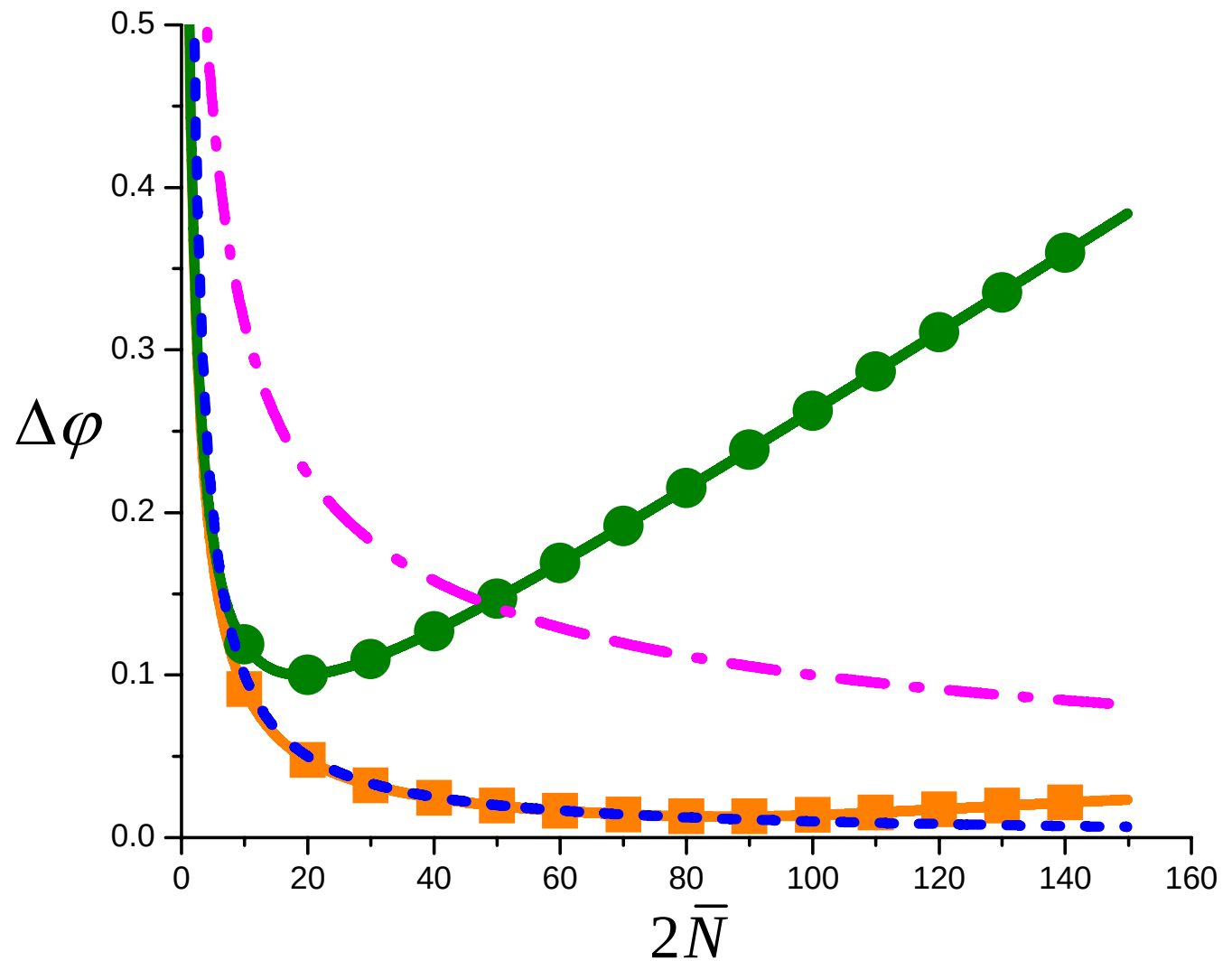

Fig. 7(a):

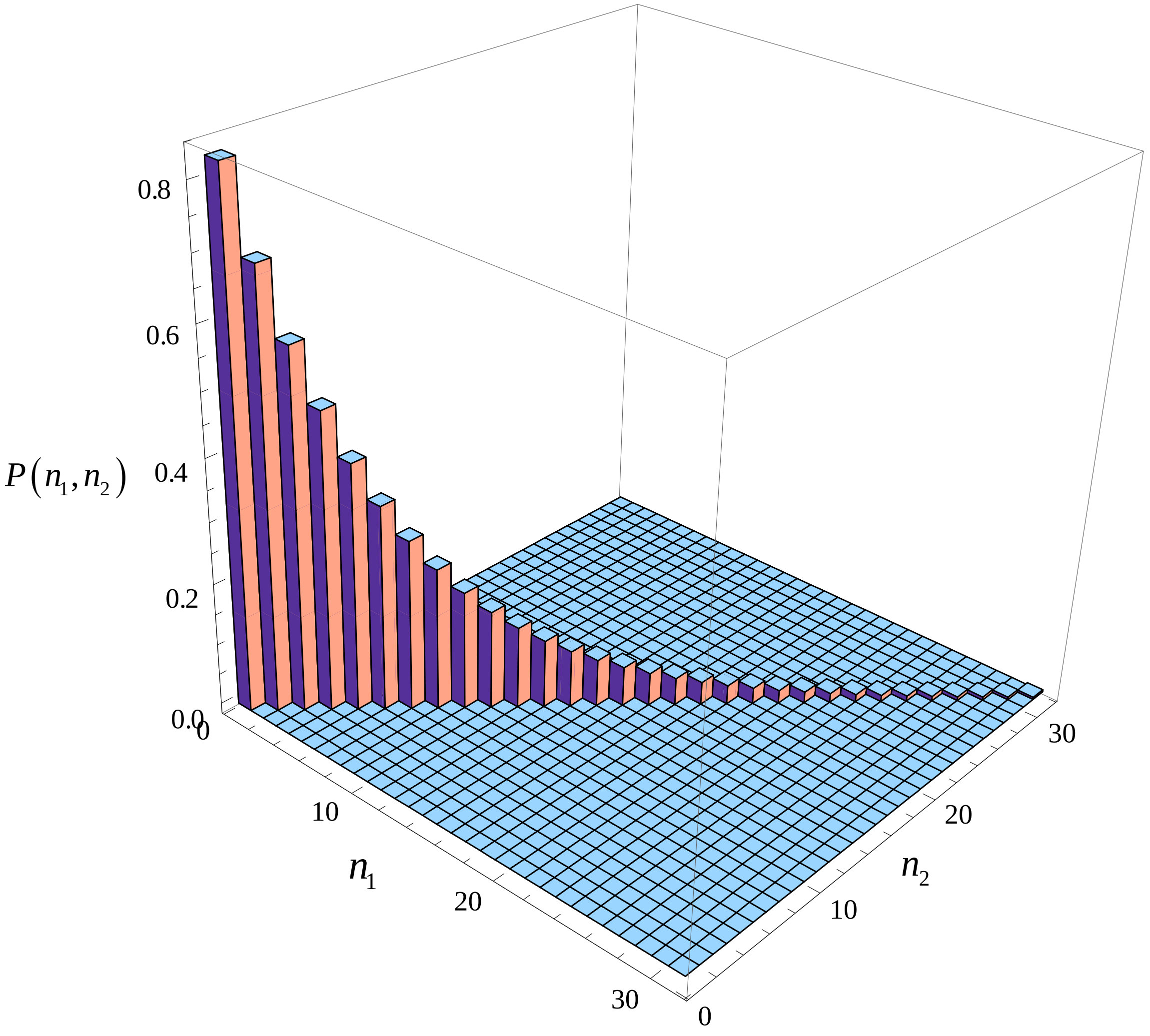

Fig. 7(b):

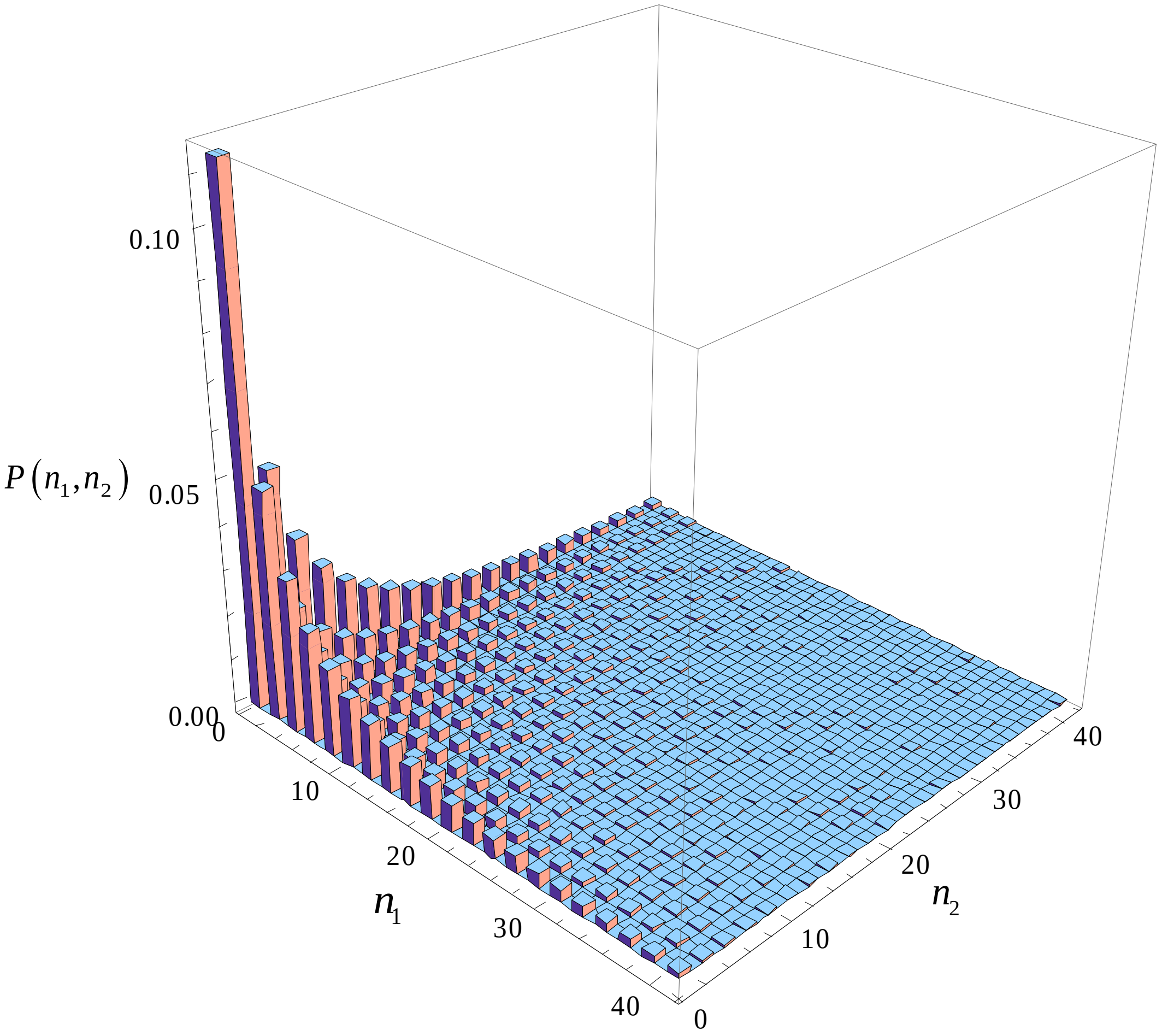

Fig. 8:

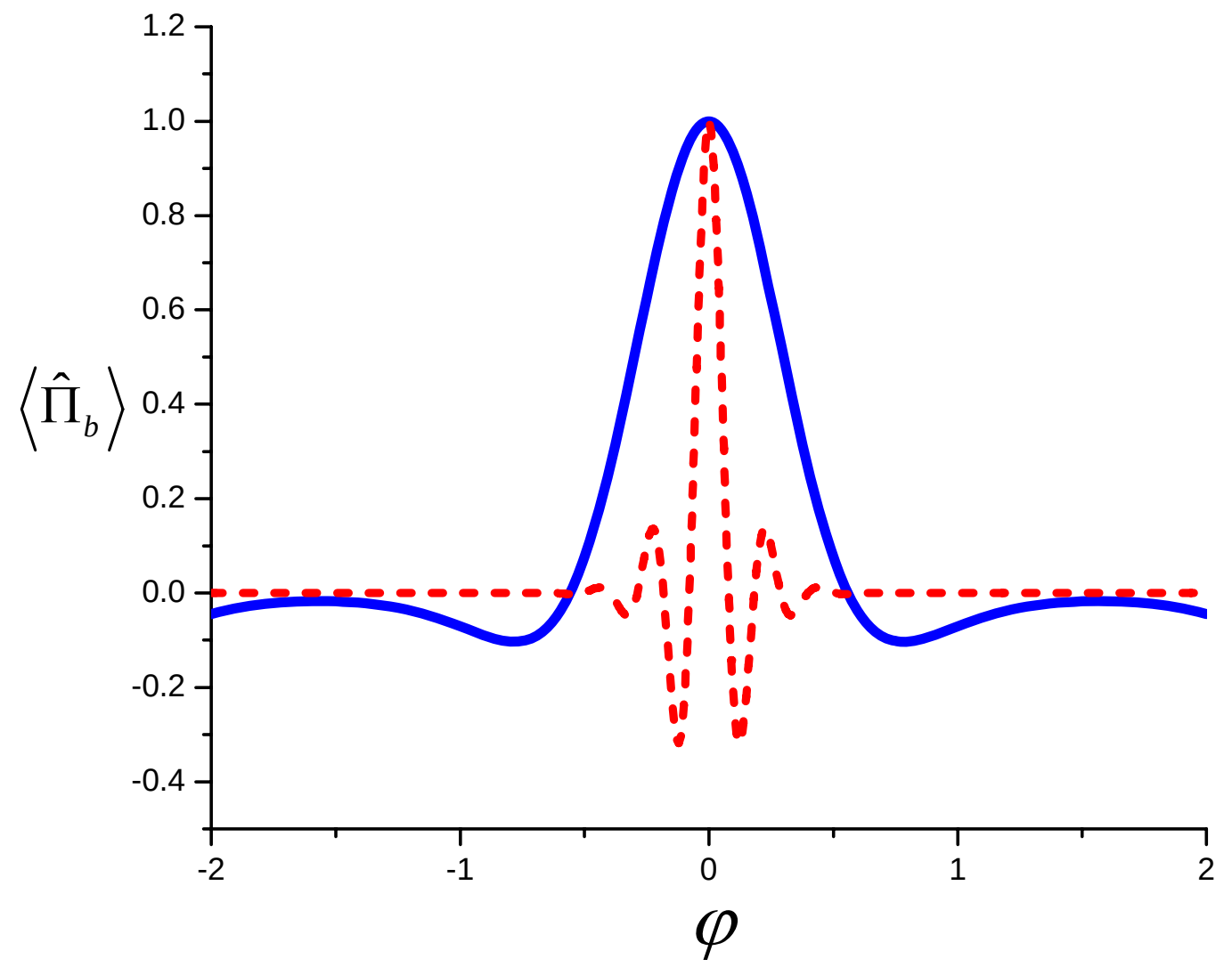

Fig. 9:

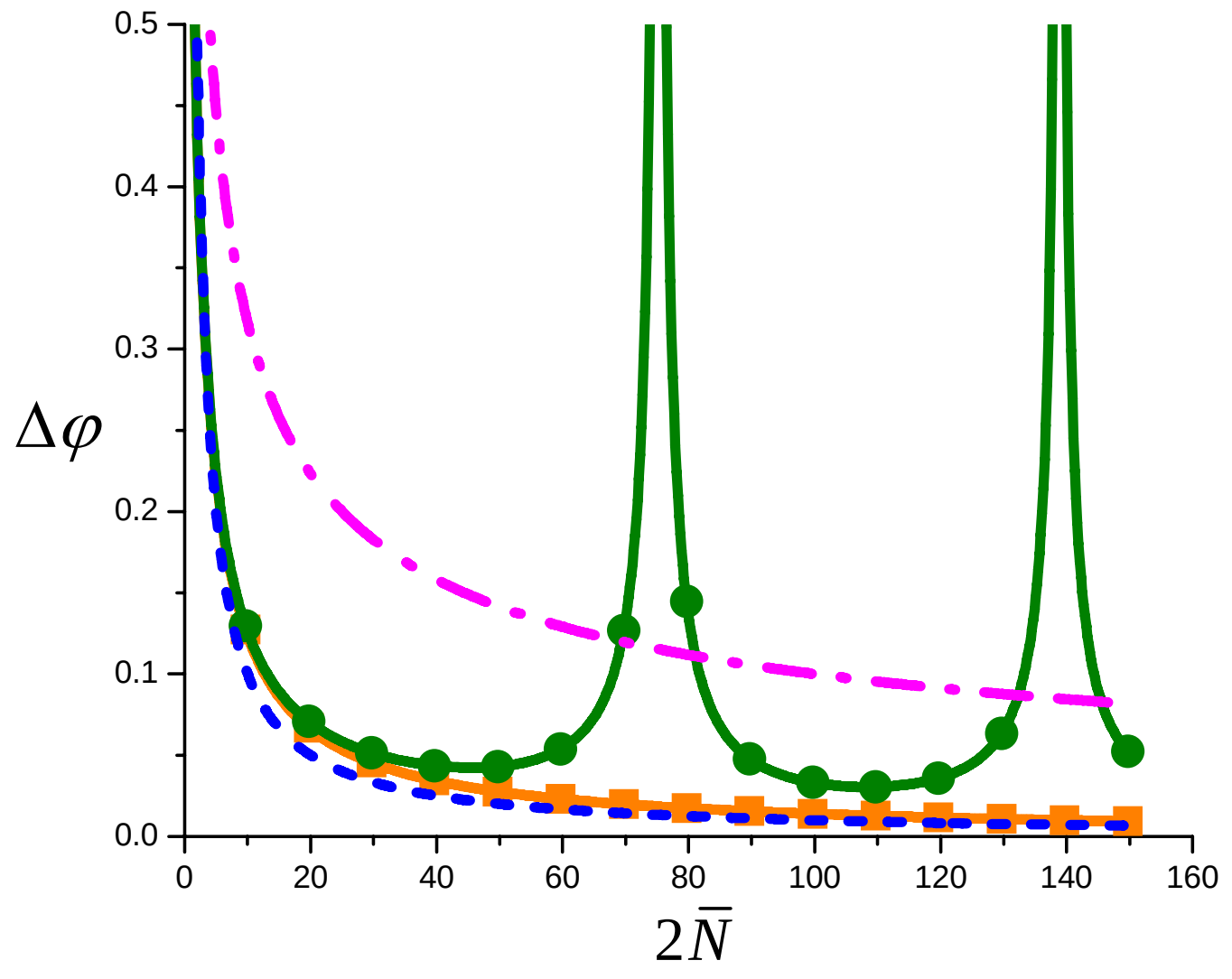

Fig. 10(a):

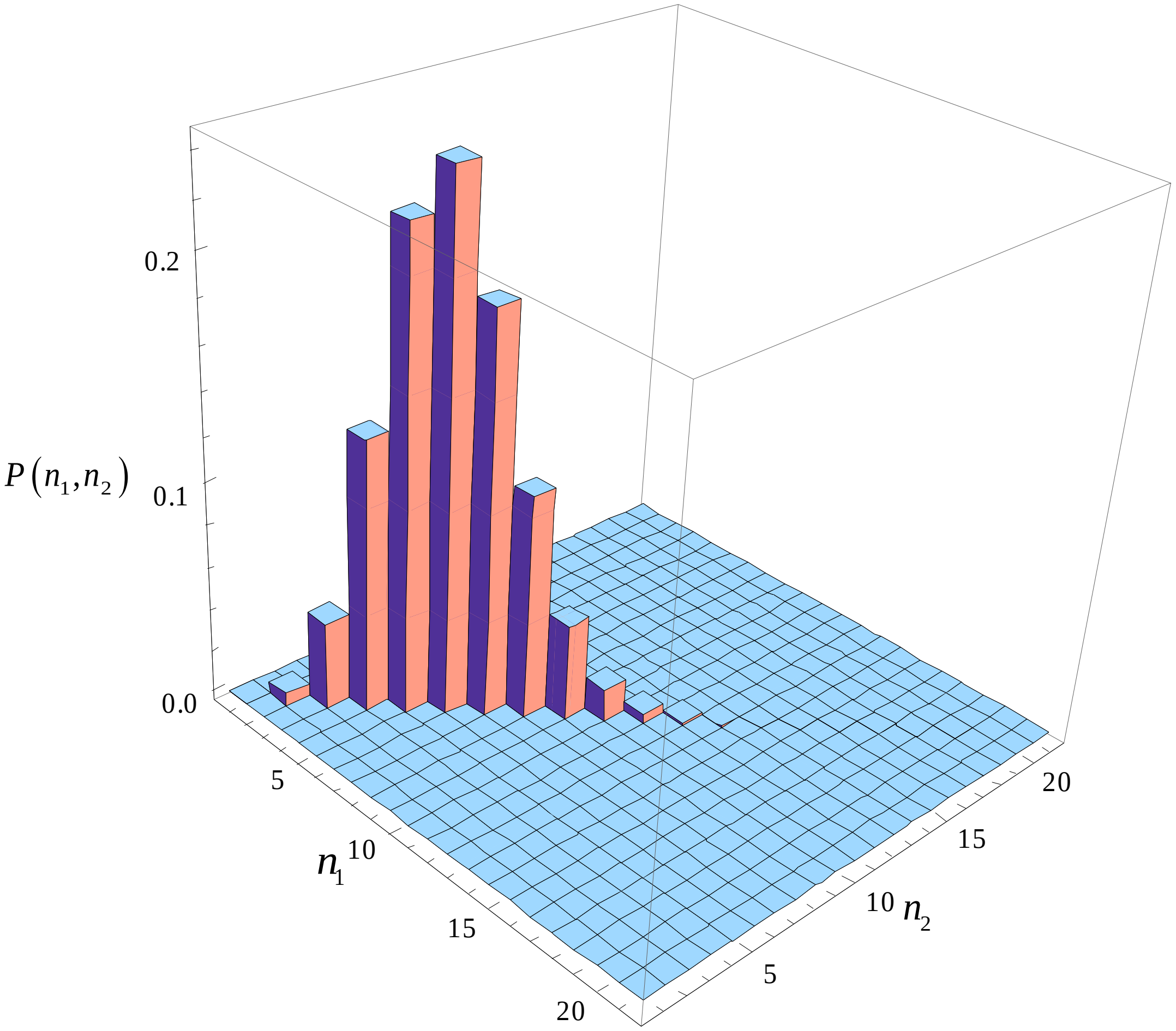

Fig. 10(b):

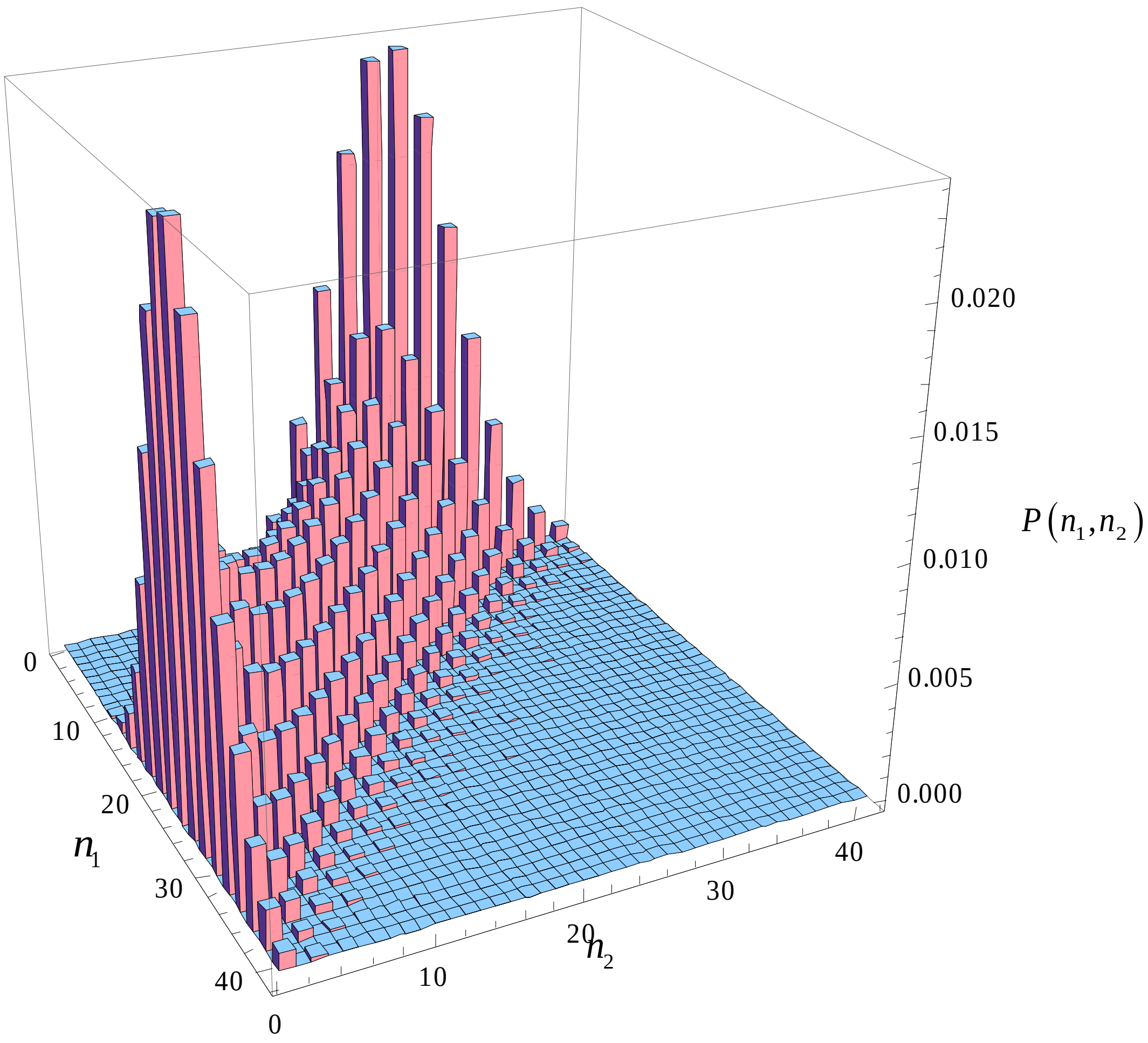

Fig. 11:

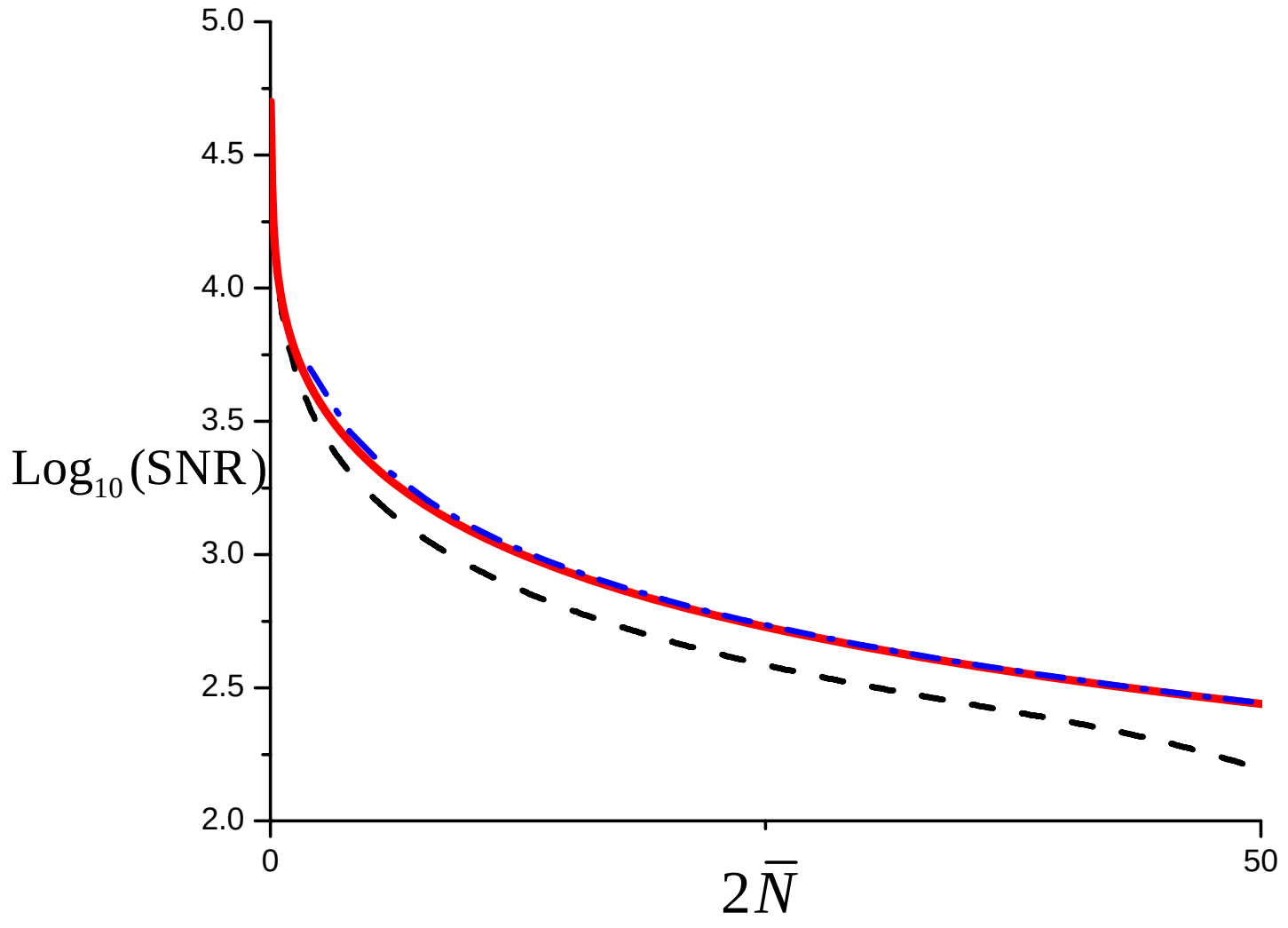